# Nodal quasiparticle in pseudogapped colossal magnetoresistive manganites

N. Mannella[1,2], W. Yang[1,2], X. J. Zhou[1,2], H. Zheng[3], J. F. Mitchell[3],
J. Zaanen[1,4], T. P. Devereaux[5], N. Nagaosa[6,7], Z. Hussain[2] & Z.-X. Shen[1]

[1]*Departments of Physics, Applied Physics, and Stanford Synchrotron Radiation Laboratory,
Stanford University, Stanford, California 94305, USA.*
[2]*Advanced Light Source, Lawrence Berkeley National Laboratory, Berkeley, California 94720, USA.*
[3]*Materials Science Division, Argonne National Laboratory, Argonne, Illinois 60439, USA.*
[4]*Instituut Lorentz for Theoretical Physics, Leiden University, POB 9506, 2300 RA Leiden, The Netherlands.*
[5]*Department of Physics, University of Waterloo, Waterloo, Ontario N2L 3G1, Canada.*
[6]*CREST, Department of Applied Physics, The University of Tokyo, 7-3-1,
Hongo, Bunkyo-ku, Tokyo 113-8656, Japan.*
[7]*Correlated Electron Research Center, AIST, Tsukuba 305-8562, Japan*

**A characteristic feature of the copper oxide high-temperature superconductors is the dichotomy between the electronic excitations along the nodal (diagonal) and antinodal (parallel to the Cu–O bonds) directions in momentum space, generally assumed to be linked to the 'd-wave' symmetry of the superconducting state. Angle-resolved photoemission measurements in the superconducting state have revealed a quasiparticle spectrum with a d-wave gap structure that exhibits a maximum along the antinodal direction and vanishes along the nodal direction[1]. Subsequent measurements have shown that, at low doping levels, this gap structure persists even in the high-temperature metallic state, although the nodal points of the superconducting state spread out in finite 'Fermi arcs'[2]. This is the so-called pseudogap phase, and it has been assumed that it is closely linked to the superconducting state, either by assigning it to fluctuating superconductivity[3] or by invoking orders which are natural competitors of d-wave superconductors[4,5]. Here we report experimental evidence that a very similar pseudogap state with a nodal–antinodal dichotomous character exists in a system that is markedly different from a superconductor: the ferromagnetic metallic groundstate of the colossal magnetoresistive bilayer manganite $La_{1.2}Sr_{1.8}Mn_2O_7$. Our findings therefore cast doubt on the assumption that the pseudogap state in the copper oxides and the nodal-antinodal dichotomy are hallmarks of the superconductivity state.**

$La_{1.2}Sr_{1.8}Mn_2O_7$ (LSMO) is a prototypical bilayer manganite that exhibits the colossal magnetoresistance (CMR) effect—the extremely large drop in resistivity induced by application of a magnetic field near the Curie temperature ($T_C$)[6,7]. The CMR effect exploits a metal–insulator transition between a low-temperature ferromagnetic-metallic ground state and a high-temperature paramagnetic-insulating phase. The nature of the ferromagnetic-metallic ground state in LSMO remains highly controversial. On the one hand, the underlying Fermi surface and band structure have clear resemblance to those of the copper oxides[8] (Lin, H., Sahrakorpi, S., Barbiellini, B. & Bansil, A., personal communication on full potential band structure and Fermi surface computations for $x = 0.4$). On the other hand, previous angle-resolved photoemission (ARPES) investigations revealed no quasiparticle peak, a suppression of spectral weight at the Fermi level ($E_F$) (the pseudogap), and an unusually light effective mass, a factor of two lighter than the calculated band structure value. This last observation is puzzling given the general expectation of strong interactions in the manganites. Moreover, the value for the in-plane conductivity calculated with the ARPES parameters is nearly one order of magnitude higher than that measured by transport[9–11].

Our ARPES experiments resolve the controversy of the low-temperature ferromagnetic-metallic groundstate of LSMO by demonstrating that its electronic structure is strikingly similar to that found in the pseudogap phase of the copper oxide high-temperature superconductors (HTSC). At 20 K, well below $T_C \approx 120$ K, our data show a small but well-defined quasiparticle peak near the $(0,0)$ to $(\pi,\pi)$ nodal direction (Fig. 1), with dispersion yielding an effective mass almost six times heavier than the band structure value (Fig. 2). The weight of the quasiparticle peak diminishes rapidly away from the nodal direction while crossing over to well-nested ghost-like Fermi-surface segments (Fig. 4). This nodal–antinodal dichotomy, in terms of anisotropy, spectral lineshapes and the nested 'ghost-like' Fermi surface away from the node, is almost identical at a phenomenological level to the one found in the underdoped $Ca_{2-x}Na_xCuO_2Cl_2$ HTSC[12].

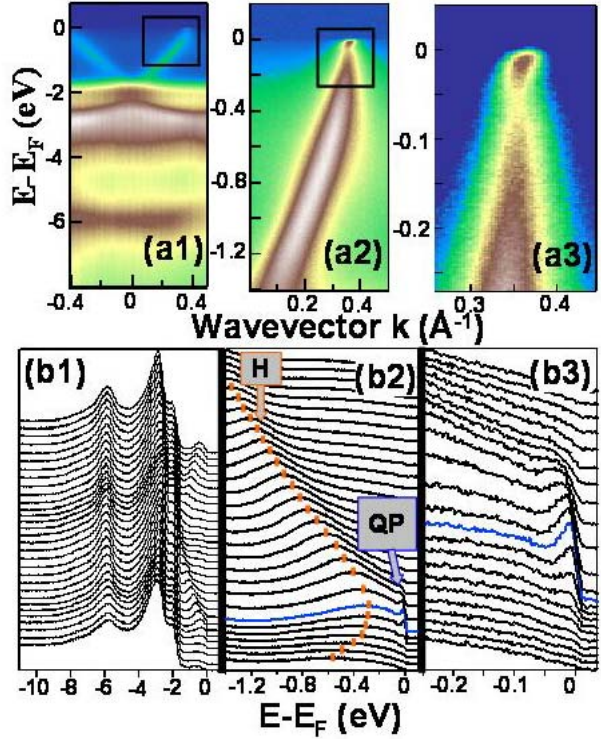

**Figure 1 Data collected along the (0,0)–(π,π) nodal direction at various magnifications. (a1)**, **(b1)**, Image plot and EDC of the valence band around the Γ point. The Mn $e_g$ states show a ≈ 1.8 eV dispersion, ≈ 0.4 eV wider than that predicted by LDA calculations[8] (Lin, H., Sahrakorpi, S., Barbiellini, B. & Bansil, A., personal communicaton). **(a2)**, Magnification of the box in **(a1)** and stacks of EDC **(b2)** of the $e_g$ band, showing the break-up of the spectral weight into a small quasiparticle peak which crosses $E_F$ at a Fermi wavevector $k_F$ ≈ 0.37 Å$^{-1}$, and a rapidly dispersing broad peak (the hump, H). **(a3)**, Magnification of the box in **(a2)** and relative EDC **(b3)**. The data have been collected at 20 K with photon energy $h\nu$ = 42 eV and the light polarization vector lying in the sample plane.

Figure 1 shows the data collected along the nodal direction at various magnifications. Because we are not dealing with a d-wave superconductor, the term 'nodal direction' does not have any special meaning in the context of manganites. Nevertheless, we use this terminology because it is useful for comparisons with the copper oxides. A striking feature is the particularly 'clean' band seen to terminate at $E_F$. This corresponds to the Mn $e_g$ states, with a bandwidth of ≈ 1.8 eV and ≈ 0.4 eV wider than the band structure prediction using a local density approximation (LDA)[8] (Lin, H., Sahrakorpi, S., Barbiellini, B. & Bansil, A., personal communication). To our knowledge, such a well-defined and wide dispersion has not previously been observed in any transition metal oxide. The sharp energy scale near 300 – 400 meV, ubiquitous in HTSC owing to antiferromagnetic interactions, is absent here[13,14].

Upon magnification, the image plots (Figs 1(a2)-(a3)) and the 'peak–dip–hump' structure in the energy distribution curves (EDC) (Figs 1(b2)-(b3)) show some of the canonical signatures of a strong coupling to bosonic modes close to the polaronic limit, albeit highly momentum-dependent[15–18]. The spectral weight is split into two branches: (1) a small but well-defined quasiparticle peak with a very narrow dispersion below ≈ 50 meV and (2) a broad 'hump' peak whose maximum disperses along a roughly parabolic band from ≈ 300 meV below $E_F$ down to the Γ point at ≈ 1.8 eV.

In Fig. 2 we compare the LDA results in ref. 8 and in the work of H. Lin, S. Sahrakorpi, B. Barbiellini & A. Bansil (personal communication) with the $e_g$ band dispersion obtained from the EDC and momentum distribution curves (MDC) analyses of the data shown in Fig. 1. The high-energy incoherent branch (the hump) starts near 300 meV and tracks the LDA dispersion. This two-branched character is reminiscent of the two poles solution for bosonic modes coupling considered in ref. 19, but the tiny intensity of the quasiparticle peak in the EDC indicates that the electron–phonon interaction is so strong here that the theory based on the conventional metallic state fails. In fact, denoting the quasiparticle effective mass by $m^*$ obtained from the low energy EDC dispersion and the bare band mass obtained from fitting the LDA dispersion by $m_{LDA}$, the value of the coupling constant $\lambda$ expressed according to the Eliashberg textbook definition as $\lambda = [(m^*/m_{LDA}) - 1]$ is found to be $\approx$ 4.6 (Fig. 2), and according to canonical theory, we would have expected the crystal to have run into a structural instability.

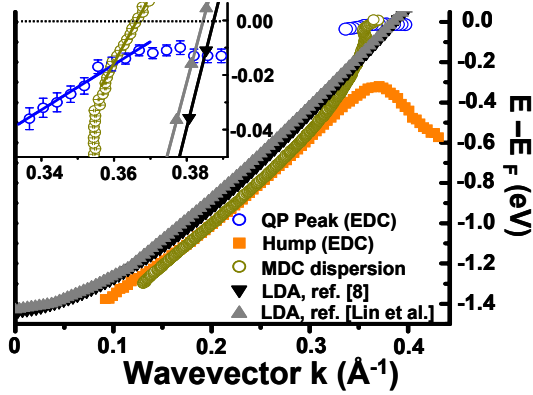

**Figure 2 Dispersion of the $e_g$ band.** Aside from a rigid shift, there seems to be a good correspondence for energies higher than $\approx$ 300 – 400 meV between the $e_g$ band dispersions calculated with LDA[8] (Lin, H., Sahrakorpi, S., Barbiellini, B. & Bansil, A., personal communication) and extracted with the MDC analysis of the data shown in Fig. 1(a2). Also shown are the positions of the broad peak (hump) and the quasiparticle peak determined from the EDC shown in Figs 1(b2) and 1(b3). The inset shows the magnification of the region close to the Fermi level crossing. The solid lines indicate the results of the fittings. If not shown, the error bars are smaller than the symbol's size. The MDC analysis for energies approaching $E_F$ leads to an S-shaped dispersion with two kinks near 50 and 100 meV. This S-shaped dispersion is an artefact resulting from the way the MDC analysis handles the Englesberg–Schrieffer two-poles solution[19]: namely the distribution of spectral weight for energies ranging from $\approx$ 50 meV to the maximum of the broad hump peak. Nevertheless, the energy scale revealed by the 50 meV kink is still meaningful, as it marks the asymptotic energy separating the coherent quasiparticle peak from the hump, as also shown in Fig. 1(a3).

Much as in HTSCs, the Fermi surface maps show well-nested straight sections along the antinodal direction (Fig. 3). As for the copper oxides, plots like those shown in Fig. 3 cannot discriminate between the true Fermi surface (sharp quasiparticle) and the ghost-like Fermi surface with no quasiparticle peaks. As indicated by the spectral lineshapes for different cuts through the Brillouin zone (Fig. 4), sharp quasiparticle peaks are only found close to the nodal points (Fig. 4(a1)-4(a3)), while the spectra near the antinodal points (Fig. 4(a6)) look quite similar to the spectra for corresponding momenta in the pseudogap state of underdoped HTSC or in a polaron state as exhibited in undoped insulating copper oxides such as $Ca_2CuO_2Cl_2$ and $La_2CuO_4$ (refs 15, 20). As in the copper oxides, the transition from well-defined nodal quasiparticle to the faint features of the segments along the antinodal directions takes place very rapidly in momentum space.

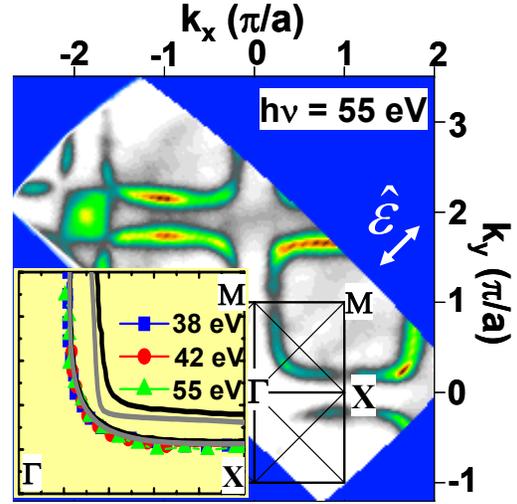

**Figure 3 Fermi surface topology.** The Fermi surface has been determined by integrating the spectral weight in a $\pm$ 50 meV window around $E_F$ of spectra excited with photons of 55 eV. Reducing the integration window to $\pm$ 10 meV results in the same spectral weight distribution. The momentum distribution of spectral weight is strongly affected by the photon energy and especially the experimental geometry, as a result of the impact that symmetry has on the matrix elements[2]. According to LDA calculations, the Fermi surface topology consists of a small electron pocket derived from $3z^2$-like states centred at the $\Gamma$ point and $x^2-y^2$-derived hole pockets centred at the M point and split into two by the hybridization between the bilayers[8] (Lin, H., Sahrakorpi, S., Barbiellini, B. & Bansil, A., personal communication). We found that (data not shown) the small electron pocket centred at $\Gamma$ is enhanced when the photon polarization vector is perpendicular to the sample plane, unlike in the cases shown here, for which the light polarization vector lies in the sample plane directed along the nodal direction (white arrow). This geometry was nonetheless used because it gives better exposure of the spectral weight at $E_F$. To better quantify the Fermi surface, we made use of the MDC analysis, which identifies as Fermi wavevectors ($k_F$) the maximum positions of the momentum distribution of spectral weight at $E_F$. The result is shown in the inset, where the Fermi surface contour extracted with the MDC analysis of data collected at three different photon energies is compared to the LDA calculations from ref. 8 (black line) and the work of Lin, H., Sahrakorpi, S., Barbiellini, B. & Bansil, A. (personal communication) (grey line). Despite an extensive search involving the use of different photon energy and/or experimental geometries, we observed only the $x^2-y^2$ band closest to the $\Gamma$ point, which tracks the LDA prediction extremely well, giving us confidence that our data are truly representative of the LSMO compound.

To illustrate the nodal–antinodal dichotomy further, we show in Fig. 4(b1) and 4(b2) the MDC spectra corresponding to the nodal and antinodal cuts denoted in Fig. 4(b3). In contrast to the nodal MDC spectra, which are sharply defined and dispersing, the antinodal MDC spectra are dispersionless for energies as high as 100 meV, a behaviour which is almost identical to that observed in $Ca_{2-x}Na_xCuO_2Cl_2$, further stressing the similarity to the nodal–antinodal dichotomy in the copper oxides[12]. We searched extensively for the presence of quasiparticle peaks with different photon energies and experimental geometries, but did not find any evidence of a quasiparticle peak other than along the nodal direction within the resolution of our experiments, leading us to conclude that the quasiparticles along the straight Fermi surface segments in the antinodal regions are suppressed below the detection limit, with Fermi arcs in the vicinity of the nodal points where discernible and well-defined quasiparticles exist.

These observations provide an interpretation of the transport measurements. Relevant to the analysis of the transport parameters is the strong mass enhancement of the quasiparticle, whose fitting yields a value for the effective mass $m^* \approx 3.3 m_e \approx 5.6 m_{LDA}$, where $m_e$ denotes the free electron mass (Fig. 2). From the MDC peak width $\Delta k$ at $E_F$ we estimate the mean free path to be $l \approx$ $1/\Delta k = 25$ Å which, when divided by the bare Fermi velocity[21], yields a scattering rate of $\tau \approx 3.28$ fs. Using these parameters in the Drude formula, we obtain an approximate value for the electrical resistivity $\rho = m^*/ne^2\tau \approx 0.66 \times 10^{-3}$ Ω cm, in fairly good agreement with the in-plane resistivity value of $\approx 2 \times 10^{-3}$ Ω cm measured by transport[22]. This result indicates that the transport properties of this compound are determined by the nodal quasiparticles and that their low mobility $\mu = e\tau/m^*$ is responsible for the relatively high value (for a metal) of the resistivity. Explaining the coexistence of metallic transport with the ghost-like straight Fermi surface segments, the presence of the nodal quasiparticles is thus a key missing element which reconciles the metallic transport properties with previous ARPES measurements in the bilayer manganites.

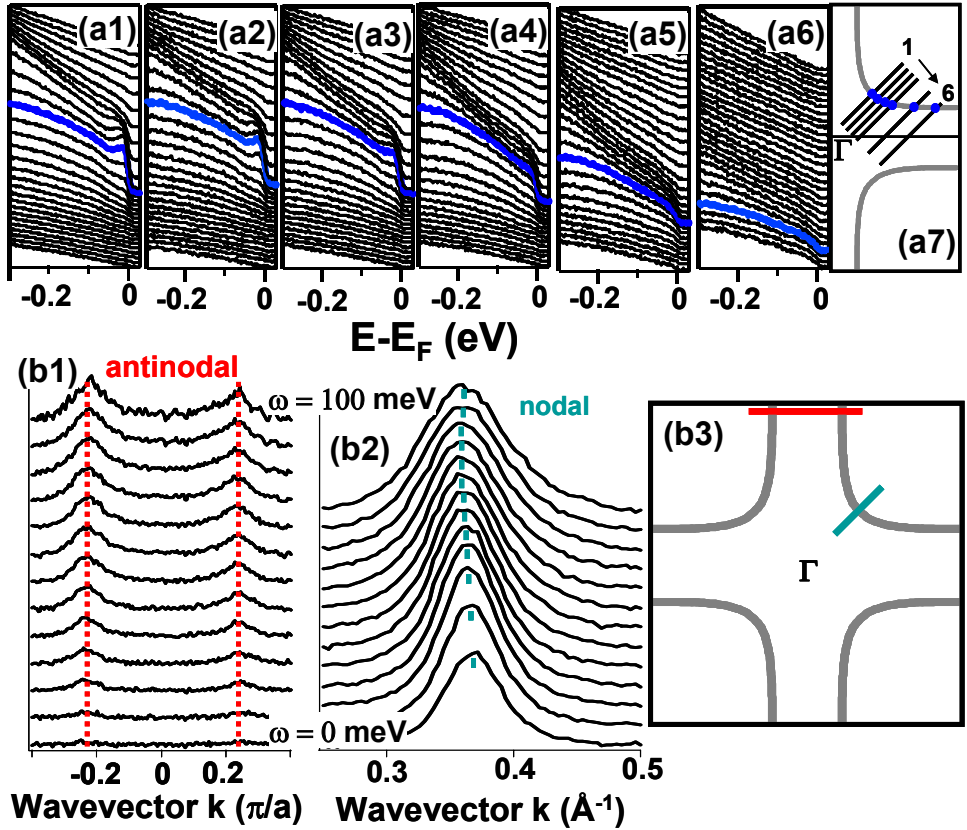

**Figure 4 Nodal–antinodal dichotomy in momentum space. (a1)-(a6)**, EDC corresponding to the cuts shown in **(a7)**. The spectral weight of the well-defined quasiparticle peak rapidly loses intensity while moving away from the nodal direction. **(b1)**, MDC spectra for the antinodal; **(b2)**, nodal cuts shown in **(b3)**. The presence of quasiparticle peaks along the nodal direction and their absence along the antinodal sections of the Fermi surface is not related to an anisotropic distribution of spectral intensity due to matrix element effects. This is evident from Fig. 3, which shows that there are regions in the parallel Fermi surface sections that contain strong spectral intensity, but do not yield well-defined quasiparticle peaks such as those along the node.

Our findings demonstrate that the ferromagnetic-metallic ground state in the bilayer manganites is far from being the highly delocalized metal described by the textbook double-exchange model[6,7]. Despite the immense differences between their electronic structures both at high energy (the large 1.8 eV 'clean' dispersion in the manganite versus the small 0.3 – 0.4 eV 'bare' bandwidth and strong damping in copper oxides due to antiferromagnetism) and at low energy (ferromagnetism versus superconductivity), underdoped copper oxides and layered manganites have a surprisingly similar mechanism in common that produces the characteristics of a nodal–antinodal dichotomous pseudogap metal. Metallic ferromagnetism and superconductivity are entirely different phenomena, so the characteristics of the pseudogap metal clearly do not have a direct connection with these ground state orders, although the underlying physics may be important for both cases. Instead, our findings indicate that the solution of the nodal–antinodal dichotomy puzzle lies elsewhere and suggest the occurrence of a phase which is ubiquitous in transition metal oxides and theoretically poorly understood: the polaronic metal with anisotropic band structure.

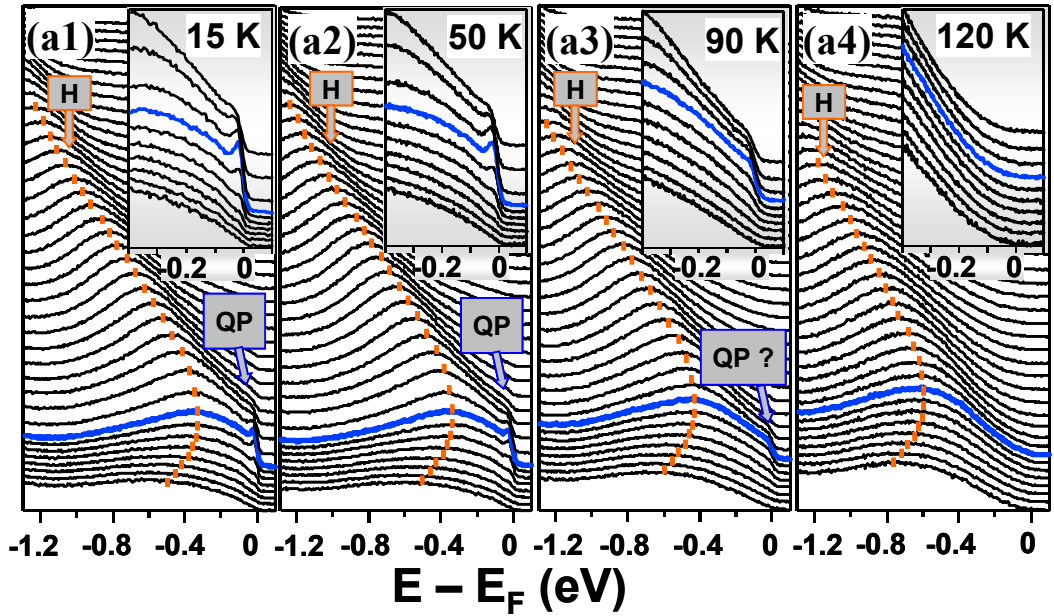

**Figure 5  Temperature-dependent evolution of the nodal quasiparticle.** The data have been collected along the $(0,0)$–$(\pi,\pi)$ nodal direction with photon energy $h\nu = 42$ eV and light polarization vector lying in the sample plane. Note the very small changes with temperature of the broad peak (hump, H) as opposed to the dramatic changes of the quasiparticle peak (insets). The quasiparticle peak - visible in **(a1)** and **(a2)** - becomes hard to track at $T = 90$ K $= 0.75\ T_C$ **(a3)** and is undetectable at the Curie temperature $T_C = 120$ K **(a4)**, which is concomitant with a metal–insulator transition between a low-temperature ferromagnetic-metallic ground state and a high-temperature paramagnetic-insulating phase. To ensure reproducibility of the data and exclude the occurrence of any surface damage during the measurements, the spectra have been collected by both increasing and decreasing the temperature.

Although the issue is not settled for the copper oxides, there is no doubt that for the manganites the electron–phonon coupling is an essential microscopic ingredient. The essence of the CMR effect is that at a temperature that is a tiny fraction of the Fermi energy the system turns from a non-degenerate liquid into a degenerate (apparently 'nodal') Fermi liquid[23]. Scattering measurements have provided direct evidence that the high-temperature classical liquid is formed from lattice polarons[24]. These form in turn a strongly correlated liquid phase, showing pronounced short-range-ordered correlations[25]. The onset of ferromagnetism tips the balance, via the double-exchange mechanism, in favour of a quantum-coherent, degenerate metal[26]. Our data support the notion that on microscopic scales this low-temperature metal is not very different from the high-temperature liquid.

Moreover, the local charge-ordered correlations observed at $T_C$ seem to have a precursor already at temperatures as low as 20 K ($T/T_C \approx 0.16$), discernible as the presence of the well-nested straight sections of the Fermi surface. For this reason, the ferromagnetic-metallic ground state may be referred to as an 'incipient' precursor to the charge-ordering state. The intriguing correlation between the nesting characteristics of the antinodal ghost-like Fermi surface and the observed charge ordering may not be coincidental. In fact, ARPES investigations of underdoped $Ca_{2-x}Na_xCuO_2Cl_2$, for which checkerboard charge ordering has been reported[27], revealed the existence of quasiparticles only along the nodal direction and the intriguing correlation between the charge-ordering periodicity and the antinodal parallel segments of the ghost-like Fermi surface[12]. To confirm this additional commonality with the copper oxides will require a systematic study of the doping dependence of the Fermi surface and charge correlations.

Our findings offer some guidance for the development of a theoretical understanding of this mysterious metallic state. We have already emphasized that the conventional metal theory (that is, the Migdal–Eliashberg theory) has non-physical ramifications when applied to the 'self-energy kinks' as shown in Figs 1 and 2. Instead, it makes more sense to invoke recent results from polaron theory[16,17]. For a single polaron the spectral function consists of a low-energy 'zero-phonon' quasiparticle peak representing the centre of mass of the quantum motion of the polaron, and a high-energy incoherent resonance. It was recently demonstrated that in general the peak position of the incoherent resonance has to track closely the bare dispersion[9,15–17]. This reflects the motions of the undressed electron confined in the polaron cloud, and a similar picture has been suggested for underdoped copper oxides[15].

In bilayer manganites, the temperature-dependent evolution of the 'peak–dip–hump' structure in the EDC (Fig. 5) offers an unprecedented opportunity to confirm that this interpretation of the high-energy dispersion shown in Figs 1 and 2 is indeed correct. As expected within this interpretation, upon increasing temperature the high-energy branch does not undergo dramatic changes, while the quasiparticle peak disappears because the polarons lose the coherence associated with the polaron centre of mass. The challenge facing theory is to explain why the centre of mass sector (quasiparticle) turns itself into a 'nodal' Fermi liquid characterized by the Fermi arcs.

**Methods**
Our experiments were performed on high-quality single crystals of composition $La_{1.2}Sr_{1.8}Mn_2O_7$ ($x$=0.4, LSMO) grown with the floating-zone method[28]. The ARPES measurements were carried out on beamline 10.0.1 at the Berkeley Advanced Light Source (ALS) using a Scienta R4000 electron analyser in the angle mode, which allows spectra to be recorded simultaneously within an angular aperture of 30° or 14°. We used the 30° mode for the Fermi surface maps and the 14° mode for the spectra with higher momentum resolution. The angular resolutions of the 30° and the 14° modes are ± 0.5° and ± 0.15°, corresponding to ≈ 0.03 Å$^{-1}$ and ≈ 8.3×10$^{-3}$ Å$^{-1}$, respectively. The samples were cleaved in ultrahigh vacuum (≈ 1–2 × 10$^{-11}$ torr) and measured at 20 K with a total instrumental energy resolution of 12 – 25 meV. The outermost 10-Å-thick bilayer in LSMO single crystals cleaved at 300K in air is insulating, with no long-range ferromagnetic order[29]. Our findings of a metallic ground state are not inconsistent with the results of ref. 29, owing to the different surface preparation methods.

**Acknowledgements** The work at the ALS and SSRL is supported by the DOE Office of Basic Energy Science, Division of Material Science, under contracts DE-FG03-01ER45929-A001 and DE-AC03-765F00515, respectively. The work at Stanford is also supported by NSF grant DMR-0304981 and ONR grant N00014-04-1-0048-P00002. The work at Argonne National Laboratory is supported by the U.S. Department of Energy Office of Science under Contract No. W-31-109-ENG-38.

**Correspondence** and requests for materials should be addressed to Z.-X.S. (zxshen@stanford.edu) or N.M. (NMannella@lbl.gov).